\documentclass[12pt]{article}
\pdfoutput=1
\usepackage{latexsym}
\usepackage[nosort]{cite}
\usepackage{epsfig,amssymb,euscript,slashed}
\usepackage[normalem]{ulem} 
\usepackage{amsmath}
\usepackage{float}
\usepackage[linktocpage]{hyperref}
\usepackage{bbm}
\usepackage{braket}
\usepackage{xcolor}
\usepackage{fancybox}
\usepackage{geometry}
\if{}
\fi

\def\be{\begin{equation}}
\def\ee{\end{equation}}
\def\bseq{\begin{subequations}}
\def\eseq{\end{subequations}}

\def\bea{\begin{eqnarray}}
\def\eea{\end{eqnarray}}

\def\bseq{\begin{subequations}}
\def\eseq{\end{subequations}}

\numberwithin{equation}{section} 
\usepackage[]{graphicx}

\def\sqr#1#2{{\vcenter{\vbox{\hrule height.#2pt
 \hbox{\vrule width.#2pt height#1pt \kern#1pt \vrule width.#2pt}\hrule
 height.#2pt}}}}

\def\slashchar#1{\setbox0=\hbox{$#1$}           
\dimen0=\wd0                                 
\setbox1=\hbox{/} \dimen1=\wd1               
\ifdim\dimen0>\dimen1                        
\rlap{\hbox to \dimen0{\hfil/\hfil}}      
#1                                        
\else                                        
\rlap{\hbox to \dimen1{\hfil$#1$\hfil}}   
/                                         
\fi}

\begin{document}
\font\cmss=cmss10 \font\cmsss=cmss10 at 7pt

\title{The Incomprehensible Simplicity of PST}

\author{ Dmitri P. Sorokin \footnote{e-mail: {\tt  dmitri.sorokin@pd.infn.it }}}

\date{}

\maketitle

\vspace{-1.5cm}

\begin{center}

\vspace{0.5cm}
\textit{I.N.F.N. Sezione di Padova, Via F. Marzolo 8, 35131 Padova, Italy\\
and
\\
Dipartimento di Fisica e Astronomia ``Galileo Galilei",  Universit\`a degli Studi di Padova
}
\end{center}

\vspace{5pt}

\abstract{ There is a hearsay that the PST approach to the covariant Lagrangian description of relativistic duality invariant theories is somewhat complicated. I will show, using an example of a chiral two-form field theory in six space-time dimensions, that it is actually natural and very simple. It directly stems from the Hamiltonian formulation of chiral-form theories.
}
\noindent


\thispagestyle{empty}


\newpage

\setcounter{footnote}{0}

\tableofcontents

\newpage

\section{Introduction}
The construction of actions for duality-symmetric or self-dual field theories in various space-time dimensions is a somewhat non-trivial problem that has been under study for more than fifty years. Historically, probably for the first time it was addressed in concrete terms in 1970 by Zwanziger \cite{Zwanziger:1970hk} who proposed a manifestly duality-symmetric Lagrangian for electrodynamics of charged electric and magnetic particles. The construction involves two dual electromagnetic vector potentials and a fixed space-time vector reminiscent of the Dirac monopole string. So the Zwanziger Lagrangian is not Lorentz-invariant, but, in the end, it produces Lorentz-covariant Maxwell's equations with electric and magnetic sources. For a review of this formulation and its further developments see e.g. \cite{Blagojevic:1985sh}. Subsequently, in the 70s-80s of the last century the issue of (the impossibility of) constructing {\it Lorentz covariant} Lagrangians for duality-symmetric fields in different space-time dimensions and for self-dual (so called) chiral $2n$-form fields in $D=4n+2$ was explored by several authors \cite{Deser:1976iy,Marcus:1982yu,Siegel:1983es,Kavalov:1986ki,Floreanini:1987as,Henneaux:1987hz,Henneaux:1988gg,Tseytlin:1990nb,Tseytlin:1990va}, especially in the context of supergravity and superstring theories. It was realized that without the use of auxiliary fields it is not possible to construct a duality-invariant Lagrangian which would be manifestly Lorentz invariant. If one accepts to abandon Lorentz covariance, then the most natural formulation for the duality-symmetric fields is the Hamiltonian one in which first-order (in time derivative) actions for these fields are non-manifestly Lorentz invariant  \cite{Deser:1976iy,Floreanini:1987as,Henneaux:1987hz,Henneaux:1988gg,Schwarz:1993vs}. In particular, for chiral $2n$-form fields in $D=4n+2$ such actions were first constructed by Henneaux and Teitelboim \cite{Henneaux:1987hz,Henneaux:1988gg} as a generalization of the two-dimensional chiral boson action of Floreanini and Jackiew \cite{Floreanini:1987as}. The variation of such an action with respect to chiral-form fields produces equations of motion whose general solution is the linear Hodge self-duality condition on the $(2n+1)$-form field strength of the chiral $2n$-form.

It is however desirable to have a manifestly Lorentz covariant Lagrangian formulation of relativistic field theories, since it drastically simplifies their coupling to gravity and other relativistic fields. To construct actions which would be simultaneously manifestly invariant under duality and Lorentz symmetry one needs to introduce auxiliary fields. Several approaches to construct such actions have been proposed in the course of time \cite{Siegel:1983es,Kavalov:1986ki,McClain:1990sx,Pasti:1995ii,Pasti:1995tn,Pasti:1996vs,Bengtsson:1996fm,Berkovits:1996tn,Belov:2006jd,Ivanov:2014nya,Sen:2015nph,Mkrtchyan:2019opf,Townsend:2019koy,Ferko:2024zth}. Comparisons and relations between some of these approaches were discussed e.g. in \cite{Mkrtchyan:2019opf,Bandos:2020hgy,Bansal:2021bis,Avetisyan:2022zza,Arvanitakis:2022bnr,Evnin:2022kqn,Bansal:2023pnr,Evnin:2023ypu,Ferko:2024zth,Janaun:2024wya,Vanichchapongjaroen:2024tkj}. Among these approaches, the PST formulation \cite{Pasti:1995ii,Pasti:1995tn,Pasti:1996vs} is the most straightforward and the most economic covariantization of the Hamiltonian-like formulation of \cite{Henneaux:1987hz,Henneaux:1988gg,Schwarz:1993vs}. To this effect it uses a single Stueckelberg-like auxiliary scalar field associated with a certain local symmetry.  In $D=4$, upon the dualization of this auxiliary scalar field into a two-form and subsequent gauge fixing local symmetry one gets the Zwanziger construction  \cite{Zwanziger:1970hk} from the PST one \cite{Maznytsia:1998xw}. The PST formulation was used to construct, for the first time, the M5-brane action of M-theory \cite{Bandos:1997ui,Aganagic:1997zq} and its NS5-brane counterpart in type IIA string theory \cite{Bandos:2000az}, the type IIB D=10 supergravity action \cite{Dall'Agata:1997ju,Dall'Agata:1998va}, a duality-symmetric $D3$-brane action \cite{Berman:1997iz,Nurmagambetov:1998gp}, actions for $D=6$ supergravities with chiral two-form (tensor) supermultiplets \cite{Dall'Agata:1997db,Riccioni:1998pj,DePol:2000re}, a duality symmetric formulation of $D=11$ supergravity and its coupling to M2- and M5-branes \cite{Bandos:1997gd}, a duality-symmetric type IIA D=10 supergravity action \cite{Bandos:2003et} and others (for more recent use of this formalism and its variants see e.g. \cite{Pasti:2009xc,Pasti:2012wv,Ko:2013dka,Ko:2016cpw,Ko:2017tgo,Buratti:2019cbm,Buratti:2019guq}). In spite of these and other achievements, there is a rumour that the PST formalism is somewhat complicated. In these notes I will show, using an example of a chiral two-form field theory in six space-time dimensions, that this formulation is actually quite natural and very simple, since it directly stems from (and is related to) the Hamiltonian formulation of this theory.

{\bf Notation and conventions}. $D=6$ Minkowski metric has mostly plus signature. Lower case Greek letters stand for six-dimensional space-time indices ($\mu,\nu,...=0,1,\ldots, 5$), which label e.g. the $D=6$ coordinates $x^\mu$. Lower case Latin letters stand for five-dimensional space indices ($i,j...=1,\ldots, 5$) and $x^\mu=(x^0,x^i)=(t,x^i)$. The antisymmetric symbol $\varepsilon_{\mu_1\ldots \mu_6}$ is defined such that $\varepsilon_{012345}=1$.

\section{Six-dimensional 2-form gauge theory}
The reason behind the choice of a chiral 2-form gauge theory in $D=6$ and not, for instance,  four-dimensional Maxwell's electrodynamics for our consideration is that in $D=4$ one has at hand the conventional Lorentz-covariant Maxwell Lagrangian which, though being not duality-invariant, produces duality-symmetric free Maxwell's equations. So, in a sense, having a manifestly duality-symmetric Lagrangian for $D=4$ electrodynamics is not indispensable. There is no such an option for the chiral 2-form gauge theory in $D=6$, and in general for the chiral $2n$-forms in $D=4n+2$, as we will see below.

\subsection{Non-chiral 2-form gauge theory}
Let us start with the consideration of a free theory of a gauge (antisymmetric) 2-form field \linebreak{$A_2=\frac 12 dx^\nu\wedge dx^\mu\,A_{\mu\nu}(x)$.} The field strength of $A_2$ is the three form $F_3=dA_2$ whose components are
\be\label{F3}
F_{\mu\nu\rho}=\partial_{\mu}A_{\nu\rho}+\partial_\nu A_{\rho\mu}+\partial_\rho A_{\mu\nu}\,.
\ee
The field strength is invariant under the gauge transformation with a vector-field parameter $\lambda_\mu(x)$
\be\label{gaugesym}
\delta A_{\mu\nu}=\partial_\mu \lambda_\nu-\partial_\nu \lambda_\mu\,.
\ee
The simplest gauge invariant Lagrangian density which one can construct with the use of $F_3$ is
\be\label{Lnonchiral}
\mathcal L=-\frac 1{3!} F_{\mu\nu\rho}F^{\mu\nu\rho}\,.
\ee
Its variation produces the free-field equation of motion
\be\label{eomF}
\partial_\mu F^{\mu\nu\rho}=0\,.
\ee
This equation is satisfied by both self-dual and anti-self-dual field strengths
\be\label{F+}
F_{\mu\nu\rho}=F^{*}_{\mu\nu\rho}\equiv \frac 1{3!}\varepsilon_{\mu\nu\rho\lambda_1\lambda_2\lambda_3}F^{\lambda_1\lambda_2\lambda_3}
\ee
and
\be\label{F-}
F_{\mu\nu\rho}=-F^{*}_{\mu\nu\rho}\,.
\ee
Therefore, the 2-form gauge field $A_2(x)$ is not chiral. Its field strength is transformed under a reducible representation of the Lorentz group $SO(1,5)$. However, in many physically interesting theories, like $D=6$ supergravities, there appear chiral 2-form gauge fields whose field strengths are either self-dual or anti-self-dual and are thus associated with an irreducible representation of $SO(1,5)$. Then the question is whether there exists a Lagrangian whose variation with respect to $A_2$ would produce only one of the self-duality conditions, e.g. \eqref{F+}. It turns out that to construct such a Lagrangian without the use of auxiliary fields one should sacrifice manifest Lorentz invariance. Instead, one might naively try to introduce the self-duality condition \eqref{F+} into the Lagrangian with the use of a Lagrange multiplier self-dual field $\Lambda_{\mu\nu\rho}=\Lambda^*_{\mu\nu\rho}$
\be\label{LLnonchiral}
\mathcal L=-\frac 1{3!} F_{\mu\nu\rho}F^{\mu\nu\rho}+\Lambda^{\mu\nu\rho}(F_{\mu\nu\rho}-F^*_{\mu\nu\rho})\,.
\ee
The variation of this Lagrangian density with respect to $\Lambda_3$ produces the desired self-duality condition. However, its variation with respect to $A_2$ produces a dynamical equation of motion of $\Lambda_3$ which implies that the theory described by \eqref{LLnonchiral} contains two chiral physical fields and not one as we wish. The Lagrangian density \eqref{LLnonchiral} is a starting point for a Lagrangian formulation of chiral p-form fields with the use of infinite number of auxiliary fields \cite{McClain:1990sx,Bengtsson:1996fm,Berkovits:1996tn}. Alternatively, it is the starting point of the formulation by Sen \cite{Sen:2015nph,Sen:2019qit} (see \cite{Barbagallo:2022kbt,Lambert:2023qgs,Hull:2023dgp,Janaun:2024wya} for latest developments and references) in which an additional chiral p-form is a ghost-like field (whose kinetic term has a wrong sign), which however completely decouples from all physical fields in the theory including gravity. We will not follow these routes, but choose instead a different one which passes through the Hamiltonian formulation \cite{Deser:1976iy,Floreanini:1987as,Henneaux:1987hz,Henneaux:1988gg,Schwarz:1993vs}.

\subsection{Hamiltonian formulation}
To get the Hamiltonian associated with the Lagrangian density \eqref{Lnonchiral} let us separate the time and space indices in the latter:
\be\label{Lnonchiral0+5}
\mathcal L=-\frac 1{3!} F_{\mu\nu\rho}F^{\mu\nu\rho}=\frac 12 F_{0ij}F_0{}^{ij}-\frac 1{3!}F_{ijk}F^{ijk}\,.
\ee
It is now convenient to introduce analogs of electric and magnetic fields
\bea\label{EB}
E_{ij}&:=&F_{0ij}=\partial_0A_{ij}-\partial_iA_{0j}+\partial_jA_{0i}\nonumber\\
B^{ij}&=&\frac 1{3!}\varepsilon^{ijklm}F_{klm}=\frac 1{2}\varepsilon^{ijklm}\partial_kA_{lm}
\eea
in terms of which the Lagrangian density takes the following form
\be\label{LnonchiralEB}
\mathcal L=\frac 12 (E_{ij}E^{ij}-B_{ij}B^{ij})\,.
\ee
Next step is to define the canonical momenta conjugate to the velocities $\partial_0 A^{\mu\nu}:=\dot  A^{\mu\nu}$. These are
\be\label{Dij}
D_{ij}=\frac{\delta \mathcal L}{\delta \dot A^{ij}}=E_{ij}
\ee
and
\be\label{D0j}
D_{0j}=\frac{\delta \mathcal L}{\delta \dot A^{0j}}=0,
\ee
since $A^{0j}$ does not enter the Lagrangian under the time derivative. So \eqref{D0j} is the primary constraint, using the Dirac terminology.

Upon the Legendre transform, the canonical Hamiltonian density is
\be\label{CH}
\mathcal H=\dot A^{ij}D_{ij}-\mathcal L=\frac 12 (D_{ij}D^{ij}+B_{ij}B^{ij})-2A_{0j}\partial_iD^{ij}\,,
\ee
where the last term was obtained by integrating by parts and skipping the total derivative (assuming appropriate boundary conditions). The field $A_{0j}$ plays the role of the Lagrange multiplier associated with the secondary constraint
\be\label{Gauss}
\partial_iD^{ij}=0,
\ee
 which is nothing but the Gauss law in this 2-form electrodynamics. The general solution of \eqref{Gauss} (at least in topologically trivial backgrounds) is
 \be\label{D=dtildeA}
 D^{ij}=\frac 1{2}\varepsilon^{ijklm}\partial_k\tilde A_{lm},
 \ee
 where $\tilde A_{lm}$ is a two-form field which is different from $A_{lm}$ in \eqref{EB}. These fields are independent physical degrees of freedom of the Hamiltonian formulation of the non-chiral two-form gauge theory.

 \subsection{First-order form of the Lagrangian density for the two-form gauge field and its splitting into chiral and anti-chiral part}

 Using the Hamiltonian density \eqref{CH} with $D_{ij}$ defined by \eqref{D=dtildeA} one gets the first-order form (in time derivative) of the Lagrangian density \eqref{Lnonchiral0+5}
 \be\label{FOL}
 \mathcal L=E^{ij}D_{ij}-\frac 12 (D_{ij}D^{ij}+B_{ij}B^{ij})\,.
 \ee
Let us now perform the following change of variables
\be\label{Changeof}
A^\pm_{ij}=\tilde A_{ij}\pm A_{ij},  \qquad B^{\pm\, ij}=\frac 12 \varepsilon^{ijklm}\partial_k A^{\pm}_{lm}=D^{ij}\pm B^{ij},\qquad E_{ij}=\frac 12 (E^+_{ij}-E^-_{ij})
\ee
with
\be\label{Epm}
E^{\pm}_{ij}=\partial_0 A^\pm_{ij}-\partial_i A^\pm_{0j}+\partial_j A^\pm_{0i}\,.
\ee
In terms of these new variables the Lagrangian density \eqref{FOL} takes the following form (modulo total derivatives)
 \be\label{FOLpm}
 \mathcal L=\frac 14(E^{+\,ij}B^+_{ij}-B^+_{ij}B^{+\,ij})-\frac 14 (E^{-\,ij}B^-_{ij}+B^-_{ij}B^{-\,ij})\,.
 \ee
The Lagrangian density splits into the sum of two terms, one of which only depends on the two-form gauge field $A^+_2$ and another one on $A^-_{2}$.

As we will now show, the first term is the Lagrangian density for the chiral two-form field $A^+_2$ whose three-form field strength is self-dual on the mass shell, while the second term is the Lagrangian density for the anti-chiral two-form field $A^-_2$ whose field strength is anti-self dual on the mass shell.

To see this, let us assume that the self-duality condition \eqref{F+} holds. By separating time and space components in \eqref{F+}, one can see that it is equivalent to
 \be\label{E=B}
 E_{ij}=B_{ij}\,.
\ee
In the Hamiltonian formulation of the free theory $E_{ij}=D_{ij}$, so for the chiral 2-form field we have the additional Hamiltonian constraint
\be\label{D=B}
 D_{ij}=B_{ij} \qquad \to \qquad \tilde A_{ij}=A_{ij}+\partial_i\lambda_j-\partial_j\lambda_i,
\ee
which implies that we now have twice less physical degrees of freedom, as desired. The above condition means that the two-form field $A^-_{2}$ defined in \eqref{Changeof} is a pure gauge and, as a consequence, $E^-_{ij}$ and $B^-_{ij}$ are zero. We are thus left with $D_{ij}=B_{ij}=\frac 12 B^{+}_{ij}$ and $E^+_{ij}$ depending on $A^+_2$ whose dynamics is described by the first term of \eqref{FOLpm}, i.e. by the Lagrangian density
\be\label{CLE}
\mathcal L_{chiral}=\frac 14(E^{+\,ij}B^+_{ij}-B^+_{ij}B^{+\,ij})\,
\ee
and the corresponding Hamiltonian density
\be\label{chiralH}
\mathcal H_{chiral}= \frac 14 B^+_{ij}B^{+\,ij}\,.
\ee
These are the Lagrangian and Hamiltonian density for the chiral two-form field \cite{Henneaux:1987hz,Henneaux:1988gg} we have been looking for, in which the magnetic field $\frac 14 B^+_{ij}$ plays the role of the canonical momentum. Analogously, the second group of terms in \eqref{FOLpm} describes the anti-chiral two-form gauge field.

We will now consider properties of the theory described by \eqref{CLE} and show that the self-duality condition \eqref{F+} indeed arises as the general solution of the equation of motion of $A^+_{ij}$ obtained by varying the Lagrangian density \eqref{CLE}. Since in what follows we will only deal with chiral-form fields, to simplify the notation we will skip the superscript $^+$ from the definition of the corresponding chiral-form quantities, which should not create any confusion.

\section{Properties of the first-order chiral 2-form field Lagrangian }

Because of the definition of $E_{ij}$  and $B_{ij}$ in eq. \eqref{EB} (and the corresponding definitions of $E^+_{ij}$ and $B^+_{ij}$ in \eqref{Changeof} and \eqref{Epm}), the components $A_{0j}$ of the chiral form potential enter the Lagrangian density \eqref{CLE} only under a total derivative. Therefore, with an appropriate choice of boundary conditions they do not contribute to the action
\be\label{Haction}
\mathcal I_{chiral}=\int d^6x \,\mathcal L_{chiral}=\frac 14\int d^6x\,(E^{ij}B_{ij}-B_{ij}B^{ij})
\ee
and to the corresponding equations of motion. This implies that in this formulation $A_{0j}(x)$ are pure gauge degrees of freedom under local shifts
\be\label{LS1}
A_{0i}(x)\to A_{0i}(x)+ \Phi_{i}(x),
\ee
which form an extra local symmetry of the action $\mathcal I_{chiral}$ in addition to the gauge symmetry \eqref{gaugesym}.

Let us now derive the equations of motion by taking the  variation of \eqref{CLE}
with respect to $A_{ij}$. The result is
\be\label{Cem}
\varepsilon^{ijklm}\partial_{k}(E_{ij}-B_{ij})=0\,.
\ee
At least in the topologically trivial situation under consideration \footnote{For the analysis of the chiral-form equations of motion in topologically non-trivial backgrounds see \cite{Bandos:2014bva}.}, the general solution of this equation is
\be\label{sdc}
E_{ij}-B_{ij}=\partial_{i}\Phi_j(x)-\partial_j\Phi_i(x)\,.
\ee
Using the local shift symmetry \eqref{LS1} we can absorb the right hand side of this equation into $E_{ij}$ thus getting the self-duality condition
\be\label{sdcc}
E_{ij}=B_{ij}
\ee
as the general solution of the equations of motion. Though \eqref{sdcc} does not have a manifestly Lorentz-invariant form, it actually is equivalent to the Lorentz-covariant Hodge self-duality condition \eqref{F+}, as we have already discussed.

Let us now consider the issue of the Lorentz invariance of the Lagrangian density \eqref{CLE}. By construction, it is manifestly invariant under the $SO(5)$ subgroup of $SO(1,5)$. So let us see how the Lagrangian transforms under the Lorentz boosts of $A_{\mu\nu}$ with a constant parameter $\lambda_0{}^k$. Setting for simplicity $A_{0j}=0$ with the use of the local symmetry \eqref{LS1}, we have
\be\label{dA}
\delta A_{ij}=x^0\lambda_{0}{}^{k}\partial_k A_{ij}+\lambda_{0k}x^k\partial_0A_{ij}\,.
\ee
Under these transformations the Lagrangian density \eqref{CLE} varies as follows (modulo total derivatives)
\be\label{dCLE}
\delta \mathcal L_{chiral}=-\frac 14\delta A_{ij}\varepsilon^{kijlm}\partial_k(E_{lm}-B_{lm})
=\frac 18\lambda_{0k}\varepsilon^{kijlm}(E_{ij}-B_{ij})(E_{lm}-B_{lm}).
\ee
 As was realized in \cite{Floreanini:1987as,Henneaux:1987hz,Henneaux:1988gg}, this variation is canceled by adding an unconventional term to the Lorentz boosts \eqref{dA} of $A_{ij}$ as follows
\bea\label{+b}
\hat\delta A_{ij}=\delta A_{ij}-\lambda_{0k}x^k(E_{ij}-B_{ij})=x^0\lambda_{0}{}^{k}\partial_k A_{ij}+\lambda_{0k}x^k B_{ij}.
\eea
Therefore, the action associated with the Lagrangian density \eqref{CLE} is invariant under the $SO(5)$ rotations and the modified (non-manifest) boosts \eqref{+b}. Note that on the mass shell \eqref{sdcc} the  transformations \eqref{+b} become the standard Lorentz boosts in compliance with the fact that the self-duality condition \eqref{F+} is manifestly Lorentz covariant.

Now we would like to promote the Lagrangian density \eqref{CLE} to a Lorentz covariant expression. This is achieved with the use of the PST formulation.

\section{PST construction of Lorentz-covariant Lagrangians for chiral forms}
The idea of how to covariantize the Lagrangian density \eqref{CLE} was prompted by so-called `Lorentz harmonics', auxiliary fields taking values in the vector representation of the Lorentz group used for twistor-like and (super)embedding descriptions of the dynamics of relativistic particles, strings and branes (see e.g. \cite{Sorokin:1999jx,Bandos:2023web} for reviews). Let us assume that in the theory under consideration there exists an auxiliary vector field $v^\mu(x)$ with the time-like unit norm \cite{Pasti:1995ii}
\be\label{v}
v_\mu v^\mu=-1
\ee
and that there is a local symmetry which allows to choose the gauge $v_\mu=-\delta^0_\mu$.\footnote{The choice of the minus sign is purely conventional. With this choice $E_{\mu\nu}$ and $B_{\mu\nu}$ reduce to $E_{ij}$ and $B_{ij}$ (rather than to $-E_{ij}$ and $-B_{ij}$) of the Hamiltonian formulation.}
We use this $v_\mu(x)$ to promote $E_{ij}$ and $B_{ij}$ to Lorentz covariant quantities
\be\label{Emn}
E_{ij} \qquad \to \qquad E_{\mu\nu}=F_{\mu\nu\rho}v^\rho
\ee
and
\be\label{Bmn}
B_{ij} \qquad \to \qquad B_{\mu\nu}=F^*_{\mu\nu\rho}v^\rho\,.
\ee
In the gauge $v_\mu= -\delta^0_\mu$ these reduce to $E_{ij}$ and $B_{ij}$, respectively.
Then we generalize eq. \eqref{Haction} as follows
\be\label{LPST}
\mathcal S_{PST}=\int d^6x \,\mathcal L_{PST}=\frac 14 \int d^6x\,(E_{\mu\nu}B^{\mu\nu}-B_{\mu\nu}B^{\mu\nu})\,.
\ee
Upon a reordering of its terms, the chiral-form Lagrangian \eqref{LPST} can be written in the following form \cite{Pasti:1996vs}:
 \be\label{LPSTa}
 \mathcal L_{PST}=-\frac 1{4\cdot 3!}F_{\mu\nu\rho}F^{\mu\nu\rho}-\frac 18 v^\mu(F_{\mu\nu\rho}-F^*_{\mu\nu\rho})(F^{\nu\rho\lambda}-F^{*\nu\rho\lambda})v_\lambda\,.
 \ee
 The simplest way to prove the equivalence of \eqref{LPST} and \eqref{LPSTa} is to use the identity
\be\label{id1}
 F^{\mu\nu\rho}=-3v^{[\mu}E^{\nu\rho]}-\frac 12 \varepsilon^{\mu\nu\rho\lambda\delta\sigma}v_\lambda B_{\delta\sigma}.
 \ee
 Note that, modulo an overall factor $\frac 14$ which can be removed by re-scaling the fields, the two Lagrangians \eqref{Lnonchiral0+5} and \eqref{LPSTa} differ by the second term in \eqref{LPSTa}, which is bi-linear in the anti-self-dual part of the field-strength $F_3$.

Now, let us assume that the local shifts \eqref{LS1} of the 2-form potential take the following form
\be\label{PST1}
\delta A_{\mu\nu}=v_{[\mu}(x) \Phi_{\nu]}(x)\,
\ee
and require that the Lagrangian density \eqref{LPST} is invariant under these transformations. One can check that this is the case if and only if
\be\label{v=d}
v_{\mu}=\frac{\partial_\mu a(x)}{\sqrt{-\partial_\rho a(x)\partial^\rho a(x)}}
\ee
with $a(x)$ being a scalar field.

Let us now look for a local symmetry of the theory which we need for being able to fix $v_\mu=-\delta_\mu^0$, or $a(x)=-x^0=-t$. For this let us consider the variation of the PST Lagrangian with respect to $A_{\mu\nu}$ and $a(x)$. Modulo total derivative terms, the result is
\bea\label{dLPST}
\delta \mathcal L_{PST}&=&-\frac 14\delta A_{\mu\nu}\varepsilon^{\mu\nu\rho\sigma\kappa\lambda}\partial_{\rho}[(E_{\sigma\kappa}-B_{\sigma\kappa})v_\lambda]\nonumber\\
&& -\frac 14\frac{\delta a(x)}{\sqrt{-(\partial a(x))^2}}\varepsilon^{\mu\nu\rho\sigma\kappa\lambda}(E_{\mu\nu}-B_{\mu\nu})
\partial_{\rho}\big[(E_{\sigma\kappa}-B_{\sigma\kappa})v_\lambda\big]\,.
\eea
A simple way to derive \eqref{dLPST} is as follows.\footnote{I am thankful to Oleg Evnin for the discussion which revealed usefulness to explain this point. The presentation is based on the Appendix 4 of the Bachelor Thesis of Ginevra Buratti \cite{Buratti:2016ahz}. For the description of the PST variation procedure in the formalism of external differential forms in any dimension see Appendix D of \cite{Bandos:2003et}.} For the variation with respect to $A_{\mu\nu}$ it is convenient to use the PST Lagrangian density in the form \eqref{LPST}. Upon the integration by parts, this straightforwardly gives the first term in \eqref{dLPST}. For the variation with respect to $a(x)$ it is more convenient to use the Lagrangian density in the form \eqref{LPSTa}. Note also that
\be\label{delta a}
\delta_a v_\mu=(\delta_{\mu}^{\nu}+v_\mu v^\nu)\frac{\partial_\nu \delta a(x)}{\sqrt{-(\partial a)^2}}\,, \qquad v^\mu\delta v_\mu\equiv 0\,.
\ee
Thus we have
\bea\label{dLa}
\delta_{a(x)}\mathcal L_{PST}& = &- \frac 14 \delta v^\mu (F_{\mu\nu\rho}-F^*_{\mu\nu\rho})(E^{\nu\rho}-B^{\nu\rho})\nonumber\\
&=& \frac 3{4}\delta v^\mu v_{[\mu}(E_{\nu\rho]}-B_{\nu\rho]})(E_{\nu\rho}-B_{\nu\rho})+\frac 18 \delta v_\mu\varepsilon^{\mu\nu\rho\delta\sigma\lambda} (E_{\nu\rho}-B_{\nu\rho})(E_{\delta\sigma}-B_{\delta\sigma})v_\lambda\nonumber\\
&=& \frac 18 \frac{\partial_\mu (\delta a(x))}{\sqrt{-\partial a)^2}}\,\varepsilon^{\mu\nu\rho\delta\sigma\lambda} (E_{\nu\rho}-B_{\nu\rho})(E_{\delta\sigma}-B_{\delta\sigma})v_\lambda\nonumber\\
&\equiv&\frac 18 {\partial_\mu (\delta a(x))}\,\varepsilon^{\mu\nu\rho\delta\sigma\lambda}\frac{ E_{\delta\sigma}-B_{\delta\sigma}}{\sqrt{-\partial a)^2}}\,\frac{E_{\nu\rho}-B_{\nu\rho}}{\sqrt{-\partial a)^2}}\,\partial_\lambda a(x)\,,
\eea
where to pass from the first to the second line we used the identity \eqref{id1} and its dual, and to pass from the second to the third line we used \eqref{delta a} and the fact that $v^\mu(E_{\mu\nu}-B_{\mu\nu})\equiv 0$. Then integrating by parts \eqref{dLa} one gets the second term in \eqref{dLPST}.

We see that the variation \eqref{dLPST} is zero if
\be\label{PST2}
\delta A_{\mu\nu}=-\frac{\varphi(x)}{\sqrt{-\partial_\rho a(x)\partial^\rho a(x)}}(E_{\mu\nu}-B_{\mu\nu}),\qquad \delta a(x)=\varphi(x)\,.
\ee
This is the local symmetry we have been looking for, since the parameter $\varphi(x)$ can always be chosen to fix $a(x)=-x^0$. In this gauge the PST formulation reduces to the Hamiltonian one discussed in the previous Section.

 It is now instructive to see what is the origin of the modified Lorentz boosts \eqref{+b}. The gauge \hbox{$a(x)+x^0=0$} is not invariant under the Lorentz boosts $\delta x^0=\lambda^{0}{}_{k}x^k$, but it is preserved by the Lorentz boosts accompanied by the local symmetry transformation \eqref{PST2} with the appropriately chosen parameter $\varphi(x)$
$$
\hat\delta (a(x)+x^0)=\varphi(x)+\lambda^{0}{}_{k}x^k=0 \qquad \to \qquad \varphi(x)=-\lambda^{0}{}_{k}x^k=\lambda_{0k}x^k\,.
$$
Then, from \eqref{PST2} one can easily see that under these combined transformations the components $A_{ij}$ of the gauge two-form transform exactly as in \eqref{+b}.

The following comment is in order. Since $a(x)$ appears in the denominator of \eqref{v=d}, the choice $a(x)=const$ for gauge fixing the local symmetry \eqref{PST2} is inadmissible.
 As explained in Section 6.1 of \cite{Ferko:2024zth} this restriction implies non-trivial conditions on the global causal structure of space-time, which must be globally hyperbolic \cite{Bernal:2004gm,bar_notes}. It is known that this space-time property is necessary for defining the Hamiltonian (ADM) formulation of General Relativity. In the PST formulation it ensures the existence of the Hamiltonian formulation for chiral $p$-forms.

 Finally, let us consider the equations of motion  which follow from \eqref{dLPST}. The equation of motion of $A_{\mu\nu}$ is
 \be\label{eomAmn}
 \varepsilon^{\mu\nu\rho\sigma\kappa\lambda}\partial_{\rho}\big[(E_{\sigma\kappa}-B_{\sigma\kappa})v_\lambda\big]=0
 \ee
 and that of $a(x)$ is
 \be\label{aeom}
 \varepsilon^{\mu\nu\rho\sigma\kappa\lambda}(E_{\mu\nu}-B_{\mu\nu})\partial_{\rho}\big[(E_{\sigma\kappa}-B_{\sigma\kappa})v_\lambda \big]=0.
 \ee
 Comparing these two equations  we see that \eqref{aeom} is identically satisfied if \eqref{eomAmn} holds. This implies that $a(x)$ is non-physical pure gauge degree of freedom, in accordance with the presence of the local symmetry \eqref{PST2}.

 Regarding the equation of motion of $A_{\mu\nu}$, its general solution (at least in topologically trivial situations) is
 \be\label{gs}
 E_{\mu\nu}-B_{\mu\nu}=\partial_\mu\Phi_\nu-\partial_\nu\Phi_\mu\,,
 \ee
 where $\Phi_\mu(x)$ is a function which must satisfy the condition $(\partial_\mu\Phi_\nu-\partial_\nu\Phi_\mu)v^\mu=0$, because of the same property  of $E_{\mu\nu}$ and $B_{\mu\nu}$ which follows from their definitions \eqref{Emn} and \eqref{Bmn}. It can be then shown that with the use of the transformations \eqref{PST1} one can set the right-hand-side of this equation to zero and thus get the Lorentz-covariant linear self-duality condition which does not depend on the auxiliary field $a(x)$
 \be\label{sdPST}
  E_{\mu\nu}=B_{\mu\nu} \qquad \Leftrightarrow \qquad  F_{\mu\nu\rho}=F^*_{\mu\nu\rho}\,.
 \ee
 To recapitulate, important properties of the Lorentz-covariant PST Lagrangian \eqref{LPST} (or \eqref{LPSTa}) which ensure that it consistently describes the dynamics of chiral $p$-form fields are the two local symmetries \eqref{PST1} and \eqref{PST2}.

 \section{Coupling to gravity and interacting chiral-form theories}
An advantage to have a covariant formulation is obvious, it allows for the straightforward coupling of chiral p-form fields to gravity and other gauge and matter fields, as well as to introduce self-interactions of the chiral-form fields. It thus drastically simplifies the construction of actions of physical interest, like that of the M-theory 5-brane \cite{Bandos:1997ui,Aganagic:1997zq}. The corresponding Hamiltonian formulations of these theories are obtained by simply imposing a gauge fixing condition in which the auxiliary field $a(x)$ is identified with a time function.

Let us consider the following generalization of action \eqref{LPST} which includes gravity and self-interactions of the chiral field
 \be\label{GPST}
\mathcal S= \int d^6x\,\sqrt{-g}\left(\frac 14 E_{\mu\nu}B^{\mu\nu}-\mathcal H(B_{\mu\nu})\right)\,,
 \ee
 where the space-time indices are contracted with a curved metric $g_{\mu\nu}(x)$, $g=\det g_{\mu\nu}$ and $\mathcal H(B_{\mu\nu})$ (that encodes self-interactions of the chiral two-form field) is a non-linear function of the components of
\be\label{Bg}
 B_{\mu\nu}= \frac{\sqrt{-g}}6\varepsilon_{\mu\nu\rho\lambda_1\lambda_2\lambda_3}F^{\lambda_1\lambda_2\lambda_3}v^\rho\,.
 \ee
 The free theory \eqref{LPST} is recovered by setting $\mathcal H=\frac 14 B_{\mu\nu}B^{\mu\nu}$.

 For consistency, it is required that the action \eqref{GPST} is invariant under the local symmetry \eqref{PST1} and a generalization of \eqref{PST2}. It can be easily seen that the action \eqref{GPST} is indeed invariant under the transformations \eqref{PST1}, since $B_{\mu\nu}$ is invariant. As far as the transfomrations \eqref{PST2} are concerned, it turns out that they are generalized as follows
 \be\label{GPST2}
\delta A_{\mu\nu}=-\frac{\varphi(x)}{\sqrt{-\partial_\rho a(x)\partial^\rho a(x)}}(E_{\mu\nu}-H_{\mu\nu}),\qquad \delta a(x)=\varphi(x)\,,
\ee
 where
 \be\label{Hmn}
 H_{\mu\nu}=2\frac{\partial \mathcal H}{\partial B^{\mu\nu}}.
 \ee
 The action \eqref{GPST} is invariant under \eqref{GPST2} provided that the function $\mathcal H(B)$ satisfies the differential condition
 \be\label{HH=BB}
 \varepsilon^{\mu\nu\rho\sigma\kappa\delta}v_\nu H_{\rho\sigma}H_{\kappa\delta}
=\varepsilon^{\mu\nu\rho\sigma\kappa\delta}v_\nu B_{\rho\sigma}B_{\kappa\delta} \, .
 \ee
 The equations of motion obtained by varying the action \eqref{GPST} with respect to $A_{\mu\nu}$ result in the non-linear generalization of the self-duality condition \eqref{sdPST} which has the following form
 \be\label{NLsd}
 E_{\mu\nu}=H_{\mu\nu}(B)\,.
 \ee
 In the gauge $\partial_\mu a(x)=-\delta^0_\mu$ the function $\mathcal H(B_{\mu\nu})$ becomes the Hamiltonian density of the non-linear theory and the condition \eqref{HH=BB} ensures that the theory is relativistic (diffeomorphism) invariant. On the other hand, if we chose the vector $v_\mu$ to be space-like instead of being time-like, then we could gauge fix $a(x)$ along a space direction, e.g. take $a(x)=x^5$. In this case the PST formulation reduces to that of Perry and Schwarz \cite{Perry:1996mk}, which is similar to the Hamiltonian formulation but with a space coordinate $x^5$ playing the role of ``time".

 The following important comment is in order. When the auxiliary field is gauge fixed, the non-linear self-duality condition \eqref{NLsd} is not manifestly relativistic covariant anymore. However, as we have already noticed in the free theory case (see the comment below eq. \eqref{+b}), since on the shell of the self-duality condition the Lorentz (or general coordinate) transformation of the gauge field $A_{\mu\nu}$ becomes the standard (covariant) one, this implies \cite{Pasti:2012wv} that it should be possible (though non-trivial) to rewrite eq. \eqref{NLsd} in a covariant form which involves only the physical field strength $F_{\mu\nu\lambda}$ and its dual $F^*_{\mu\nu\lambda}$. In the case of the M5-brane (briefly described in the next Section) the relation between the Perry-Schwarz and the PST self-duality condition  on the one side and a covariant self-duality condition which arises in the superembedding description of the M5-brane \cite{Howe:1996yn,Howe:1997fb} (or in a closely related construction of \cite{Cederwall:1997gg}) was established in \cite{Howe:1997vn,Bandos:1997gm}. For a generic non-linear duality-invariant $p$-form theory in $D=2(p+1)$ a covariant non-linear self-duality condition was derived in \cite{Avetisyan:2021heg,Avetisyan:2022zza} in the following suggestive form
 \be\label{sdc1}
 F-F^*=\frac{\partial V(F+F^*)}{\partial (F+F^*)}\,,
 \ee
 where $V(F+F^*)$ is a generic scalar function of the self-dual part of the $F_{p+1}$ field strength.\footnote{An important idea that non-linear interactions in Lagrangians of duality-invariant theories can be described by an arbitrary function $V(\Lambda^+)$ of an auxiliary self-dual tensor field $\Lambda^+$  (which is not subject to any additional condition like \eqref{HH=BB}) was put forward and realised in four-dimensional duality-symmetric non-linear electrodynamics in \cite{Ivanov:2002ab,Ivanov:2003uj,Ivanov:2014nya}. In \cite{Avetisyan:2021heg,Avetisyan:2022zza} this idea was resurrected in D=4 and extended to higher-dimensional duality-invariant and chiral-form theories within the approach of Mkrtchyan \cite{Mkrtchyan:2019opf} based on a different set of auxiliary fields. Eq. \eqref{sdc1} was obtained in the latter approach upon the elimination of the auxiliary fields in equations of motion thereof. Later on, in \cite{Ferko:2024zth} it was shown that \eqref{sdc1} also arises as a consequence of equations of motion in the (chiral-form generalization of the) Ivanov-Nurmagambetov-Zupnik construction \cite{Ivanov:2014nya}.} In $D=6$ the relation between \eqref{sdc1} and the self-duality conditions of the Perry-Schwarz formulation was established in \cite{Avetisyan:2022zza} by a dimensional reduction to $D=5$.  Differential equations relating the PST $\mathcal H(B)$ to $V(F+F^*)$ and vice versa were obtained in \cite{Ferko:2024zth}.

 For more details on the properties of generic non-linear chiral form theories in the PST formulation see \cite{Buratti:2019guq,Bandos:2020hgy,Ferko:2024zth}.

 \subsection{M5-brane}
 An example of $\mathcal H(B_{\mu\nu})$ that satisfies \eqref{HH=BB} is the Born-Infeld-like function that appears in the M5-brane action \cite{Perry:1996mk,Pasti:1997gx}
 \be\label{BIM}
 \mathcal H_{M5}=T\sqrt{-\det\left(g_{\mu\nu}+\frac 1{\sqrt T}B_{\mu\nu}\right)}\,,
 \ee
 where $T$ is associated with the $M5$-brane tension.

 To promote \eqref{GPST} with $\mathcal H(B)$ defined in \eqref{BIM} to the full M5-brane action in eleven-dimensional supergravity \cite{Bandos:1997ui,Aganagic:1997zq} one should consider an embedding of the M5-brane worldvolume parametrized by $x^\mu$ into an $N=1$, $D=11$ superspace parametrized by supercoordinates $Z^M(x)=\left(X^m(x), \Theta^\alpha(x)\right)$, where $X^m$ are eleven bosonic coordinates and $\Theta^\alpha$ are 32 fermionic ones. One then takes the superfields of $D=11$ supergravity which are the vector supervielbein $E^a(Z)=dZ^ME_M{}^a(Z)$ $(a=0,1,\ldots,10)$ whose leading component is the $11D$ graviton, the spinor supervielbein $E^\alpha(Z)=dZ^ME_M{}^\alpha(Z)$ whose leading component is the gravitino, the three-form gauge superfield and its six-form dual
$$
C_3(Z)=\frac 1{3!}dZ^MdZ^NdZ^PC_{PNM}(Z),\qquad C_6(Z)=\frac 1{6!}dZ^{M_1}\ldots dZ^{M_6}C_{M_6\ldots M_1}(Z),
$$
 and pull them back on the M5-brane worldvolume regarding $Z^M(x)$ to be the functions of $x^\mu$.

The six-dimensional worldvolume metric $g_{\mu\nu}(x)$ is induced by the embedding and has the following form
\be\label{imetric}
g_{\mu\nu}(x)=\left(\partial_\mu Z^ME_M{}^a\right)\,\left(\partial_\nu Z^ME_M{}^b\right)\,\eta_{ab}\,.
\ee
where $\eta_{ab}$ is the eleven-dimensional tangent space Minkowski metric.

Next, one generalizes the chiral-form field strength (and corresponding fields $E_{\mu\nu}$ and $B_{\mu\nu}$ defined in \eqref{Emn} and \eqref{Bg}) by subtracting from it the pullback of $C_3(Z)$
\be\label{calF}
\mathcal F_{\mu\nu\rho}=F_{\mu\nu\rho}-\sqrt{T}\partial_\mu Z^M\partial_\nu Z^N\partial_\rho Z^PC_{PNM}(Z(x))\quad \to \quad \mathcal E_{\mu\nu}=\mathcal F_{\mu\nu\rho}v^\rho, \qquad \mathcal B_{\mu\nu}=\mathcal F^*_{\mu\nu\rho}v^\rho\,.
\ee
Thus the M5-brane action in a generic superbackground of $D=11$ supergravity is
\bea\label{SM5}
\mathcal S_{M5}&=& \int_{\mathcal M_6} d^6x\,\sqrt{-g}\left(\frac 14 \mathcal E_{\mu\nu}\mathcal B^{\mu\nu}-T\sqrt{-\det\left(g_{\mu\nu}+\frac 1{\sqrt T}\,\mathcal B_{\mu\nu}\right)}\right)\nonumber\\
&&+\frac T2\int_{\mathcal M_6} \left(C_6(Z(x))+\frac 1{\sqrt T} A_2(x)\wedge C_3(Z(x))\right)\,,
\eea
where the second integral describes minimal couplings of the M5-brane to the dual $C_6$ and $C_3$ gauge fields of $D=11$ supergravity.

The action is invariant under the local symmetries \eqref{PST1} and \eqref{GPST2} (in which $E_{\mu\nu}$ and $B_{\mu\nu}$ are replaced with $\mathcal E_{\mu\nu}$ and $\mathcal B_{\mu\nu}$), and under a fermionic $\kappa$-symmetry, a crucial ingredient of all superbrane actions (see \cite{Bandos:1997ui,Aganagic:1997zq} for more details on the structure of the M5-brane action and \cite{Bergshoeff:1998vx} for its Hamiltonian).

\subsection{Non-linear conformal chiral two-form electrodynamics}
A more recent peculiar example of the non-linear chiral two-form theory is the six-dimensional counterpart \cite{Bandos:2020hgy} of the four-dimensional ModMax electrodynamics \cite{Bandos:2020jsw}. This is the unique (classical) conformally invariant theory of the self-interacting chiral two-form gauge field in $D=6$ which contains the free theory \eqref{LPST} as a zero coupling limit. For this theory the function $\mathcal H(B_{\mu\nu})$ in \eqref{GPST}, satisfying the condition \eqref{HH=BB}, has the following form
\be\label{MMH}
\mathcal H_{MM}=\cosh(\gamma)\,s-\sinh(\gamma)\,\sqrt{s^2-p^2}\,,
\ee
where
\bea
\label{ps}
s&=&\frac 14 B_{\mu\nu}B^{\mu\nu}\,,\nonumber\\
p^{\mu} &=&-\frac 1{8\sqrt{-g}}\varepsilon^{\mu\nu\rho\lambda\sigma\kappa}B_{\rho\lambda}B_{\sigma\kappa}v_{\nu}\,, \qquad p=\sqrt{p_\mu p^\mu}\,
\eea
and $\gamma$ is a dimensionless parameter. Note that for $\gamma=0$ the function $\mathcal H_{MM}$ reduces to that of the free chiral two-form.

The fact that the action \eqref{GPST} with the function $\mathcal H_{MM}$ in \eqref{MMH} is conformally invariant can be checked by noticing that this action is invariant under the local Weyl rescaling of the metric $g_{\mu\nu}\to \Omega(x)g_{\mu\nu}$, a criterium for the theory to be conformally invariant in Minkowski space due to Zumino \cite{Zumino:1970tu}.

The $6D$ ModMax theory and its Born-Infeld-like generalization have been studied in the context of $T\bar T$-like stress-tensor deformations in \cite{Ferko:2024zth}. Exact pp-wave and (black) string solutions of this theory coupled to gravity have recently been considered in \cite{Deger:2024jfh}.

\section{Conclusion}
In these notes I have described reasoning which led to the construction of the covariant PST formulation of classical chiral $p$-form theories starting from their Hamiltonian formulation and tried to elucidate the main features of this approach. For more details about this formulation in various dimensions, its variants, applications and relation to other Lagrangian formulations of chiral $p$-forms see papers mentioned in the Introduction and references therein. Important subtle issues of the quantisation of the chiral-form field theories are beyond the scope of these notes.

\section*{Acknowledgements}
The author is thankful to Chris Ferko, Kurt Lechner and an anonymous referee for having carefully read and spotted a number of typos in this manuscript, and for the encouragement to put it on the arXiv.
The author thanks the Department of Physics, UWA for kind hospitality during his visit on Jan 8-21, 2024 under the Australian Research Council Project No. DP230101629, and acknowledges the kind hospitality and financial support extended to him at the MATRIX Program ``New Deformations of Quantum Field and Gravity Theories'' between 22 Jan and 2 Feb 2024. This work was also partially supported by the Spanish AEI MCIN and FEDER (ERDF EU) under grant PID2021-125700NB-C21 and by the Basque Government Grant IT1628-22.


\if{}
\bibliographystyle{abe}
\bibliography{references}{}
\fi

\providecommand{\href}[2]{#2}\begingroup\raggedright\endgroup

\end{document}